\documentclass[11pt]{article}

\usepackage{graphicx}

\def\be{\begin{equation}}
\def\ee{\end{equation}}

\begin{document}
\begin{titlepage}
\vspace{4cm}

\begin{center}{\Large \bf General Reaction-Diffusion Processes\\
 With Separable Equations for Correlation Functions}\\
\vspace{1cm} \centerline{\bf ${\rm V.\ \ Karimipour}$ \vspace{1cm}
\footnote{email: vahid@sharif.edu}} {\it Department of Physics,
Sharif University of
Technology,\\ P.O.Box 11365-9161,\\ Tehran,Iran} \\
\end{center}
\vskip 3cm
\begin{abstract}
We consider general multi-species models of reaction diffusion
processes and obtain a set of constraints on the rates which give
rise to closed systems of equations for correlation functions.
Our results are valid in any dimension and on any type of
lattice. We also show that under these conditions the evolution
equations for two point functions at different times are also
closed. As an example we introduce a class of two species models
which may be useful for the description of voting processes or
the spreading of epidemics.
\end{abstract}
\end{titlepage}

\section{Introduction}
An interesting class of non-equilibrium problems with a rich
dynamical behaviour and a vast area for applications are
stochastic reaction-diffusion systems (see \cite{Schutz},
\cite{hinrichsen}, \cite{spohn}, \cite{ligget} and references
therein). These are the processes in which one or several species
of particles hop randomly on a lattice, and interact in various
possible ways with each other. In the one species case, it is
convenient to denote a particle by the symbol $1$ and a vacant
site (a hole) by the symbol $0$. Then a simple hopping is
represented by $ 1 + 0 \rightarrow 0 + 1 $. In addition to
exclusion which means that no two particles can occupy a single
site, the possible interactions include pair creation ( $ 0 + 0
\rightarrow 1 + 1 $), pair annihilation ( $ 1 + 1 \rightarrow 0 +
0 $), coagulation ( $ 1 + 1 \rightarrow 1 + 0 $),  de-coagulation,
  ($ 0 + 1 \rightarrow 1 + 1 $), birth ( $0\rightarrow 1$) and
death ($1\rightarrow 0$) processes. Obviously the variety of
elementary processes rapidly increases with the number of
species.\\ In general such lattice systems are difficult to treat
by rigorous analytical means and correspondingly, considering the
vast amount of such models, relatively few exact results are
known.\\ Over the past few years the application of operator
formalism to these stochastic processes and their mapping to
quantum spin systems and their generalizations has turned to be
quite fruitful. In view of this correspondence, many of the
techniques of quantum spin systems such as free fermion
techniques, Bethe ansatz and related algebraic techniques
\cite{bet1, bet2, bet3, bet4, bet5, bet6, bet7}, have been used
in the investigation of reaction diffusion systems, specially in
one dimensional lattice systems. Also by using the operator
formalism, some of the old techniques such as the matrix product
ansatz \cite{hakim} have been put to very fruitful use, in
solution of one dimensional stochastic systems \cite{dehp} (see
\cite{Schutz, hinrichsen, derrida} and references therein).
Almost all of the above methods have one limitation, they are
restricted to one dimensional lattices, specially
if we are interested in exact solutions.\\
A common feature of any model of interacting particles and indeed
the main source of difficulty in obtaining exact solutions is that
the equations of motion of correlation functions form an infinite
hierarchy, that is, the equation of motion of $n-$point functions
include $(n+1)-$point functions and in general higher correlation
functions. One can truncate the hierarchy at a level by various
kinds of approximations, the simplest and the most common method
is to break the hierarchy at the first level by the mean field
approximation. However in some models and in low space
dimensions, the amount of diffusive mixing may not be enough to
warrant such an approximation. Some of these models may also have
long relaxation times so that their simulation may be difficult
and time consuming.  For these models exact solutions are highly
desirable.\\ Interestingly enough, there are models in which the
hierarchy of equations of correlation functions automatically
breaks at every level, hence the possibility of obtaining exact
solutions. For these models the equations of motion of $n-$point
functions depend only on $k-$point functions with $ k\leq n $. We
should stress that while a powerful technique like matrix product
ansatz or its dynamical version \cite{dynamicmpa} only transform
the problem to another equally difficult and yet convenient
problem, that is, finding the representations and calculating
matrix elements of strings of operators of an algebra
\cite{dehp,k1,kk}, for these models the simplification appears to
be genuinely effective. Moreover this property is independent of
the dimension and the geometry of the lattice. In view of its
generality this is a great simplification and deserves to
be pursued further from various directions.\\
The observation of this phenomenon in some models like symmetric
exclusion and partial exclusion processes  \cite{ligget, sp, ss1,
ss2} led G. M. Sch\"{u}tz \cite{Schutzstat} to raise the question
of classification of such models, i.e., a general criterion on the
reaction and diffusion rates such that the resulting equations
for the correlation functions decouple. By considering the one
species processes and the particle density correlation functions
$\langle n(x)\rangle$, he found that from among the $12-$
parameter family of single species reaction diffusion systems, a
$10-$ parameter family fall within this class.\\
Since then this question has been pursued further. For example
one can consider the hole-density correlation functions $\langle
1-n(x)\rangle$ (the so called empty interval method) with two
\cite{ben, aka} or three site interactions \cite{hh} or even more
general functions like $\langle a + b n(x)\rangle$ \cite{doering,
ben, peschel}. These correlation functions lead to
different sets of constraints on the rates.\\
\subsection{The aim and the results of the paper}
The aim of the present paper is to investigate this question for
the general multi-species case. That is, we assume that there are
$ p+1 $ species of particles labeled as type $ 0, 1, 2, \cdots p
$ which hop and interact in a lattice of arbitrary geometry. We
interpret particles of type $0$ as holes and other particles as
real particles.   We assume two body interactions between the
particles, and obtain the general condition on the rates, so that
the correlation functions of densities of real particles, i.e.,
particles of type $ 1, 2, \cdots p $ decouple from the
correlation functions of higher
levels in the hierarchy.\\
We show that under this condition the equation of motion of
two-time two-point correlation functions are also closed. These
types of correlation functions are important in the analysis of
voting processes \cite{vote1}.\\
In view of the arbitrary number of species in our model, one will
have much more freedom to find an exactly solvable model for
description of physical phenomena, specially in the area of
chemical kinetics, where one usually has more than one species of
particles.\\ Since in our analysis and hence in our results,
there is an asymmetry between a particle of type $0$ and other
particles, it is important to keep in mind that the
interpretation of particle $0$ as a hole is not essential.
Therefore in adapting a model from the class discussed below to a
situation of physical interest, one can change this
interpretation and take any other particle of type $1, 2, \cdots
p $ to stand for a hole and
particle $0$ for a real particle.\\
For general $p-$ species models, the number of independent rates
is $ (p+1)^4 - (p+1)^2 $. We show that the family of exactly
solvable models (in the above sense) live on an $N_p$ dimensional
manifold (hyperplane) where $ N_p := (p+1)^4-(p+1)^2 - 2p^3 $. For
the one species case we find $ N_1 = 16-4-2 = 10 $, in accord
with the results of Sch\"{u}tz \cite{Schutzstat}. For the two and
three species cases we have respectively $ N_2 = 56 $ and $ N_3 =
186$, where the number of independent rates for these models are
originally $ 72 $ and $ 240 $ respectively. For an elaboration
on this see the remark $3$ in the text.\\
As an example and for concreteness we study a $2-$ species family
and further constrain it with certain extra symmetries. In this
class we can find models suitable for the description of the
spread of an epidemic, the exchange of ideas
and votes, and the spreading of news or rumor. \\
We set up the solution of one point functions for these models
and investigate to a certain extent the properties of these
solutions.\\ The structure of this paper is as follows: In
section $2$ we introduce our notations and conventions. In
section $3$ we obtain the general conditions on the rates. In
section $4$ we consider the case $ p=2$, and by imposing further
symmetry requirement on these models we introduce a class of
solvable two-species models. In section $5$ we set up the general
solution of the one point functions for this model. We conclude
the paper with a discussion.\\

\section{Notations and conventions} Throughout the paper
we will adhere to notations which we will
collect here for convenience.\\
{\it{For ease of notations we will consider a one dimensional
lattice which may be infinite or periodic. However all our
results are valid also on lattices of arbitrary shape and
arbitrary dimension.  We will also work explicitly with nearest
neighbor interactions, although again our results are valid for
arbitrary range of interactions. These facts have already been
shown in \cite{Schutzstat} and are also easily verified by
reviewing our method of proof, in the sense that no part of the
reasoning depends on the underlying lattice or the range of
interactions.}}\\
We denote the points of the lattice by Latin letters from the end
of the alphabet ${x-1}, {x}, {x+1}, \cdots$. In a finite lattice
we number the sites from $ 1 $ to $ L $. To each site $ x$ of the
lattice, we assign a random variable $ \tau(x) $ which can take
$p+1$ values $ 0, 1, \cdots p$. We denote the values of this
random variable when it takes all the values including possibly
the value $0$, by Greek letters $ \alpha, \beta, \mu, \cdots $,
and when it takes only the values different from $0$, by Latin
letters, from the middle of the alphabet, like $ i, j, k, l
\cdot$. Thus $ \langle n_i(x)\rangle :=  \langle \delta_{\tau(x),
i}\rangle $ denotes the average density of particles of type $i$
at site $x$, or the probability of site $ x$ being occupied by a
particle of type $i$, and $ \langle n_0(x)\rangle :=  \langle
\delta_{\tau(x), 0}\rangle $ denotes the probability of this site
being empty. It thus follows that
\begin{equation}\label{sum(n)}
  \sum_{\mu=0}^{p} <n_{\mu}(x)> = 1.
\end{equation}
We assume that two particles of type $\alpha$ and $\beta$ on two
adjacent sites may transform stochastically to particles of type
$ \mu $ and $\nu$ with rate $ R_{\alpha, \beta}^{\mu, \nu}$. This
is written as:
\begin{equation}\label{rates}
\alpha , \beta \longrightarrow \mu , \nu \hskip 1cm {\rm with\ \
rate\ \ \ \ } R_{\alpha \beta}^{\mu \nu}.
\end{equation}
Since a Greek index includes also the value $ 0 $, which we
interpret as a vacant site, the above transformations include all
the possible processes conceivable for all types of particles on
the two sites. For example $ R_{i0}^{0i}$ is the hopping rate of
a particle of type $i$ and $R_{00}^{k0} $ is the rate of creation
of a particle of type $k$ from the vacuum and $R_{ij}^{k0}$ is
the rate with which two particles of type $i$ and $j$ interact or
coagulate to form a
particle of type $k$. \\
We will use the operator formalism for Markov processes. This
formalism is well known by now, since in the past few years it
has been used extensively for the analysis of reaction diffusion
processes, specially in one dimension.\\ In this formalism we
should assign a complex $p+1$ dimensional Hilbert space $C^{p+1}$
to each site, with ortho-normal basis states
\begin{equation}\label{hilbertsapce}
|0\rangle,  |1\rangle,  \cdots  |p\rangle , \hskip 2cm \langle
\mu | \nu \rangle = \delta_{\mu, \nu}.
\end{equation}
The Hilbert space of the whole lattice is the tensor product of
all the local Hilbert spaces of the sites. At any given time $ t
$, each configuration of the lattice is given by the values of
the random variables of all sites $ \tau(1), \tau(2), \cdots
\tau(L) $. Such a configuration occurs with probability $
P(\tau(1), \tau(2), \cdots \tau(L); t) = \langle \tau(1), \tau(2),
\cdots \tau(L)| P(t)\rangle $. The state vector $ |P(t)\rangle $
determines all the probabilities and its evolution is governed by
a Hamiltonian derived from the rates:
\begin{equation}\label{shrodinger}
  \frac{d}{dt}|P\rangle = H |P\rangle
\end{equation}
For convenience we have absorbed the minus sign, which is usually
included in this equation, into the definition of the Hamiltonian.
The probabilities are normalized by requiring that
\begin{equation}\label{norm}
\sum_{\tau(1), \tau(2), \cdots \tau(L)}\langle \tau(1), \tau(2),
\cdots \tau(L) | P\rangle = 1
\end{equation}
This equation can be rewritten as
\begin{equation}\label{snorm}
\langle S|P\rangle := \langle s |^{\otimes L} |P\rangle = 1
\end{equation}
where $ \langle s | $ is defined as:
\begin{equation}\label{s}
\langle s | := \sum_{\mu = 0}^{p} \langle \mu |
\end{equation}
Note that the Bra state $ \langle S|$ is the sum of all the
possible configurations of the system.\\  From the master
equation (\ref{shrodinger}) and the property $ \langle S| H = 0 $
(as required by conservation of probability), one obtains the
Heisenberg-like equation of motion for the average of any time
independent observable $ O $:
\begin{equation}\label{heis}
\frac{d}{dt}\langle O(t)\rangle = \langle S|[\hat{O},
H]|P(t)\rangle
\end{equation}
where $ \hat{O}$ is an appropriately chosen operator whose matrix
element gives the average of the observable, i.e:
\begin{equation} \langle O(t) \rangle = \langle S| \hat{O}|P(t)\rangle.
\end{equation}
Note that the time dependence of the average comes from the
evolution of probabilities, therefore a better notation for the
average of a time independent observable will be $ \langle
O\rangle (t)$. However we use the notation $\langle O(t)\rangle$,
since it will be convenient when we consider two point functions
at un-equal times. Consider a completely general Markov system
whose configurations are labeled by $ C$ and a time independent
observable $ O $ of the configurations. We have:
\begin{equation}\label{2time}
  \langle O(t_2)O(t_1)\rangle := \sum_{C_2, C_1}
  O(C_2)O(C_1)P(C_2, t_2; C_1, t_1)
\end{equation}
In the operator formalism, it is easily shown that this two point
function is represented as:
\begin{equation}\label{2timeoperator}
  \langle O(t_2)O(t_1)\rangle := \langle S| \hat{O}(t_2)\hat{O}(t_1)|P(0)\rangle
  \end{equation}
where $\hat{O}(t)$ is the Heisenberg-like operator, $
\hat{O}(t):= e^{-tH}\hat{O}e^{tH}$. The above correspondence is
also true for different kinds of operators. In particular we
have: $ \langle O(t)O(0)\rangle := \langle S|
e^{-tH}\hat{O}e^{tH}\hat{O}|P(0)\rangle $. We then have:
\begin{equation}\label{twotimeequation}
\frac{d}{dt}\langle O(t)O(0)\rangle = \langle S|[\hat{O}(t),
H]\hat{O}|P(0)\rangle = \langle S|[\hat{O},
H]e^{tH}\hat{O}|P(0)\rangle
\end{equation}
where we have used $ \langle S| e^{-tH} = \langle S|$.\\
Returning to our model, the local operators $ E_{\alpha,\beta}:=
|\alpha \rangle \langle \beta|$ act on the states of a site as:
$E_{\alpha,\beta} | \mu \rangle = \delta_{\beta, \mu}
|\alpha\rangle $. Of particular interest are the diagonal
operators $ E_{00}, E_{11}, \cdots E_{pp}$ which act as number
operators for the holes, particles of type $1$, to particles of
type $p$. Hereafter we abbreviate these diagonal operators and
show them only by one index instead of two, i.e: $ E_{\mu} $
stands for $ E_{\mu\mu}$:
\begin{equation}\label{Ei}
E_0 = \left(
\begin{array}{ccccc}
  1 &  &  &  &  \\
   & 0 &  & &  \\
   &  & 0 & & \\
   &  &  & 0 & \\
   &  & &  & 0
\end{array}
 \right)\hskip 1cm \cdots \hskip 1cm E_p = \left(
\begin{array}{ccccc}
  0 &  &  &  &  \\
   & 0 &  & &  \\
   &  & 0 & & \\
   &  &  & 0 & \\
   &  & &  & 1
\end{array}
 \right)
\end{equation}
The above operators have the commutation relations:
\begin{equation}\label{commut}
[E_{\alpha \beta},E_{\mu \nu}] = \delta_{\beta \mu}E_{\alpha \nu}
- \delta_{\nu \alpha} E_{\mu \beta}.
\end{equation}
Moreover the following property of these operators is also
important in the sequel:
\begin{equation}\label{sE}
  \langle s| E_{\alpha \beta} =  \langle s| E_{\beta \beta} =:
  \langle s| E_{\beta},
  \end{equation}
where we have abbreviated $ E_{\beta\beta} $ as $ E_{\beta}$.\\
The Hamiltonian describing the processes (\ref{rates}) is
constructed as a sum of local Hamiltonians acting on adjacent
sites:
\begin{equation}\label{Hamiltonian}
  H = \sum _x h (x, x+1)
\end{equation}
where the operator $h(x, x+1)$ means that the operator $ h $ acts
only nontrivially on sites $ x $ and $ x+1$. The operator $ h$ is
constructed from local operators as follows, where we will use
hereafter the Einstein summation convention over Greek indices:
\begin{equation}\label{localh}
h = R_{\alpha \beta}^{\mu \nu}\big (E_{\mu \alpha}\otimes E_{\nu
\beta}- E_{\alpha}\otimes E_{\beta}\big)
\end{equation}
The conservation of probability constrains the rates to satisfy
the relation
\begin{equation}\label{conserv}
\sum_{\mu, \nu} R_{\alpha, \beta}^{ \mu, \nu} = 0 \hskip 1cm
\forall \ \ \alpha ,\ \ \beta
\end{equation}
Finally we need two matrices constructed from the matrix of rates
which will prove useful later, and we prefer to introduce them
here for convenience. Out of the matrix of rates, we form a set of
matrices $R^i $ and $ S^i$ defined as follows:
\begin{eqnarray}\label{RS}
  (R^i)_{\alpha, \beta}&:=& \sum_{\nu} R_{\alpha \beta}^{i \nu}\\
  (S^i)_{\alpha, \beta}&:=& \sum_{\nu} R_{\alpha \beta}^{\nu i}
\end{eqnarray}
The conditions for decoupling of equations will be expressed in
terms of these matrices.
\section{Decoupling  of correlation functions}
The simplest correlation functions are one-point functions which
determine the average densities of particles of each type at each
site. Thus we are interested in the equation of motion of the
one-point function  $ \langle n_i(x)\rangle := \langle
\delta_{\tau(x),i}\rangle $. This is in fact the probability that
site $x$ is occupied with a particle of type $i$.  In the operator
formalism this one-point function is written as a matrix element
of the corresponding operator, namely:
\begin{equation}\label{ni(x)}
  \langle n_i(x,t)\rangle = \langle S|E_{i}(x)|P(t)\rangle =:
  \langle E_{i}(x)\rangle,
\end{equation}
where by the last equality we have defined the bracket or the
average of an operator. From equation (\ref{heis}), we find
\begin{equation}\label{ni(x)equation}
\frac{d}{dt}\langle n_i(x)\rangle =  \langle [ E_{i}(x),
H]\rangle = \langle [E_{i} (x), h (x-1,x)]\rangle +\langle [E_i
(x), h (x,x+1)]\rangle .
\end{equation}
Each of these two terms leads in general to two-point functions
on their relevant sites, the first term on $(x-1, x) $ and the
second on $(x,x+1)$. We now ask under what condition a
cancellation occurs in {\it{each}} of these terms
{\it{separately}} so that we end up with only one point functions
on the right hand site. A little reflection on the words written
in italic shows that the question we are asking and henceforth
its answer, does not depend at all on the underlying lattice and
the range of the interaction, in so far as we are only
considering Hamiltonians
with two body interactions.\\
To find the answer to our question we calculate one of these
terms say the second one. We note that:
\begin{equation}\label{eih1}
\langle S| [E_{i}(x), h(x,x+1)] = R_{\alpha \beta}^{ \mu, \nu}
\langle S| [ E_{ii}(x), E_{\mu \alpha}(x) E_{\nu \beta}(x+1) ]
\end{equation}
where we have ignored the diagonal part of $h(x,x+1)$ which
obviously commute with $ E_{ii}(x)$. Using the commutation
relations (\ref{commut}) and also the property (\ref{sE}) we find:
\begin{equation}
\langle S| [E_{ii}(x), h(x, x+1)] = \sum_{\nu} R_{\alpha
\beta}^{i \nu}\langle S| E_{\alpha}(x) E_{\beta}(x+1) -
\sum_{\mu, \nu} R_{i \beta}^{\mu \nu} \langle S| E_{i}(x)E_{\beta}
(x+1).
\end{equation}
The second term vanishes in view of equation (\ref{conserv}) and
we are left with:
\begin{equation}
\langle S| [E_{ii}, h(x, x+1)] = ({\bf R^i})_{\alpha, \beta}
\langle S|
 E_{\alpha}(x) E_{\beta}(x+1)
\end{equation}
Expanding the right hand terms we find
\begin{eqnarray}\label{eih1'}
\langle S|[E_{ii},h(x, x+1)]&=&{\bf R^i}_{0,0}\langle S|
E_{0}(x)E_{0}(x+1)+{\bf R^i}_{j,0}\langle S|E_{j}(x)E_{0}(x+1)
\cr &+& {\bf R^i}_{0,k}\langle S| E_{0}(x)E_{k}(x+1)+{\bf
R^i}_{j,k}\langle S|E_{j}(x)E_{k}(x+1),
\end{eqnarray}
where now we are using the summation convention on Latin indices
and ${\bf R^i}_{j,k} $ stands for $({\bf R^i})_{j,k}$. We now use
the operator identity
\begin{equation} E_{0} = 1 - E_{1}-
E_{2} - \cdots E_{p}
\end{equation}
and eliminate the operators $ E_{0}(x)$ and $E_{0}(x+1)$ on the
right hand side of the above equation and demand that all the
quadratic terms vanish. It is easily seen that this cancellation
occurs when the matrices $ {\bf R^{i}}$ satisfy the relations:
\begin{equation}\label{Ri}
{\bf R^i}_{j,k} = {\bf R^i}_{0,k} + {\bf R^i}_{j,0} - {\bf
R^i}_{0,0} \hskip 2cm \forall \ i,\ j,\ k.
\end{equation}
This simply means that for each matrix ${\bf R^i}$, all the
elements are fixed once the elements of the first row and column
are determined.\\ Similar calculations for the first term of
(\ref{ni(x)equation}) leads to the following condition:
\begin{equation}\label{Si}
{\bf S^i}_{j,k} = {\bf S^i}_{0,k} + {\bf S^i}_{j,0} - {\bf
S^i}_{0,0} \hskip 2cm \forall \ i,\ j,\ k.
\end{equation}
where the matrices $ {\bf S^i}$ have already been defined in
(\ref{RS}).\\
Multiplying both sides of (\ref{eih1'}) by $ |P(t)\rangle $ it is
seen that once the above conditions are satisfied, the equations
of motion of $1-$ point functions depend only on one point
functions. A short calculation shows that once the equations of
motion for one point functions are closed, it guarantees that the
equation of higher order correlation functions are also closed,
that is their equation of motion depend solely on the $k$-point
functions with $ k\leq n$. In this way the hierarchy of equations
of $n-$ point functions is terminated and closed at any level and
the system amends itself to exact analytical treatment. Every
model whose rates satisfy the relations (\ref{Ri}, \ref{Si}) is
solvable in the above sense.\\
Moreover we show that the equations of motion of two-point
functions at different times are also closed under the above
conditions. In some models like voting models \cite{vote1} the
calculation of such correlation functions are
important.\\
To see this consider the two point {\it{two time}} correlation
function $ \langle n_{i}(x,t)n_j(x',0)\rangle $. We have:
\begin{eqnarray}\label{twopoint}
 \frac{d}{dt}\langle n_{i}(x,t)n_j(x',0)\rangle &\equiv& \langle S|
 e^{-tH} [E_{ii}(x), H]e^{tH} E_{jj}(x')|P(0)\rangle \\ &\equiv& \langle S|
  [E_{ii}(x), H]e^{tH} E_{jj}(x')|P(0)\rangle
\end{eqnarray}
Under the conditions (\ref{Ri} and \ref{Si}), we know that the
commutator in the above equation is expressible as the sum of
local site operators. Restoring the operator $ e^{-tH} $, i.e., $
\langle S| \longrightarrow \langle S|e^{-tH} $, we find that on
the right hand side of this equation only two point functions
appear. Therefore the equation of motion of two-time two-point
functions will also be closed.\\
\textbf{Remarks:}\\
\textbf{1:} The emergence of two sets of constraint for the rates
does have nothing to do with a given site $ x $, having two
neighbors $ x-1 $ and $x+1$ and hence on the underlying lattice.
On a general lattice we still have only these two equations. It
only reflects the fact that the interaction Hamiltonians defined
on each link may not be symmetric under the interchange of its
ends. Therefore each site $x$ contributes to two types of
interaction Hamiltonians, namely those of the type $ h(x,y)$
which for all different $y$'s lead to (\ref{Ri}) and those of the
type $h(z,x)$ which for all different $z$'s lead to (\ref{Si}). In
fact for symmetric models in which there is no driving force one
expects that these two sets of constraint become identical. This
is in fact the case, since for these symmetric models, one has: $
R_{\alpha \beta}^{\mu \nu} = R_{\beta \alpha}^{\nu \mu}$ and
hence $ {\bf R^i}_{\alpha,\beta} =  {\bf S^i}_{\beta,\alpha}$,
which makes the two set of conditions identical.\\
\textbf{2:} For general models the number of independent rates is
$ (p+1)^4 - (p+1)^2 $. To count the number of conditions on the
rates we note that a matrix such as $ {\bf R^1} $ imposes $ p^2$
linear equations on the rates, since its first row and column
determine all the other elements. There are $2p$ such matrices
and hence the number of conditions is $2p^3$. Therefore the
number of parameters of the solvable family we are considering is
$ N_p = (p+1)^4-(p+1)^2 - 2p^3 $. For the one species case we
find $ N_1 = 16-4-2 = 10 $. For the two and three species models
we have respectively $56$ and $ 186$ free parameters
respectively.\\
\textbf{3:} It appears that the family of integrable multi-species
models live on a manifold of huge dimension and one is at
complete ease to choose many models of his own choice for
adapting them to any physical situation. However this is an
illusion and as we will see, in choosing physically interesting
models from this manifold one is much more restricted than it
appears at first sight. The reason is that we are usually
interested in simple subclasses of these models, ones in which we
can set many of the irrelevant or uninteresting parameters equal
to zero either on physical grounds or to make our analysis simple
and transparent. However it often happens that once we set some of
these parameters equal to zero, the set of admissible parameters
collapses drastically so that we are left with totally
uninteresting models or models which are actually equivalent to
one species models. Geometrically the constraints are a set of
hyperplanes which pass through the origin. The admissible rates
lie in the intersection of all these hyperplanes. However all the
rates are also constrained to be positive. It may happen that
this intersection of hyperplanes, although a manifold of high
dimension, may intersect the positive sector of the space of
parameters in a very low dimensional submanifold. As an example
consider a model with say 6 rates $ r_1, r_2, r_3, r_4, r_5 $ and
$r_6$ and one constraint $ r_1+r_2+r_3+r_4+r_5-r_6 = 0$. The set
of admissible rates is obviously a five dimensional plane.
However if for some reason we are interested in those models in
which the rate $r_6$ is vanishing, then we are left with the
constraint $ r_1 +r_2 +r_3+r_4+r_5 = 0 $ which forces all these
remaining parameters to vanish. In the multi-species case this
difficulty shows up more severely since we are dealing with a
large number of hyperplanes. Thus it is nontrivial to find
physically
interesting multi-species models in the class discussed above.\\
We will conclude this section with the final
form of the equation of motion of one point functions.\\
Collecting the remaining linear terms in the equation
(\ref{eih1'}) and its counterpart, we find (with summation
convention understood for Latin indices):
\begin{eqnarray}\label{linear1}
\frac{d}{dt} \langle n_i(x)\rangle &=& {\bf R^i}_{0,0} \langle
n_{0}(x)+ n_{0}(x+1) - 1 \rangle\cr &+&{\bf R^i}_{j,0}\langle
n_{j}(x)\rangle + {\bf R^i}_{0,j}
 \langle n_{j}(x+1)\rangle\
 \cr  &+& (x\rightarrow x-1, {\bf R}
 \rightarrow {\bf S})
\end{eqnarray}
This equation is specific to a one dimensional lattice. In a
general lattice it should be modified appropriately, the
modification is however straightforward (see section 4). Before
going to the consideration of a two-species model, it is
instructive to recapitulate the findings for the one species
model from this general perspective.
\subsection{The one species separable models}
If there is only one species of particles on the lattice, we have
only two matrices namely $ {\bf R^1}$ and ${\bf S^1} $. These are
two by two matrices, subject to the conditions:
\begin{eqnarray}\label{RS-1}
{\bf R^1}_{1,1} + {\bf R^1}_{0,0} &=& {\bf R^1}_{1,0} + {\bf
R^1}_{0,1} \\ {\bf S^1}_{1,1} + {\bf S^1}_{0,0} &=& {\bf
S^1}_{1,0} + {\bf S^1}_{0,1}
\end{eqnarray}
When expanded by using the definition (\ref{Ri}, \ref{Si}) of the
matrices $ {\bf R^1} $ and $ {\bf S^1}$, they yield:
\begin{eqnarray}\label{RS-1-expanded}
R^{10}_{11} + R^{11}_{11} + R^{10}_{00} + R^{11}_{00} &=&
R^{10}_{01} + R^{11}_{01} + R^{10}_{10} + R^{11}_{10}\\
R^{01}_{11} + R^{11}_{11} + R^{01}_{00} + R^{11}_{00} &=&
R^{01}_{01} + R^{11}_{01} + R^{01}_{10} + R^{11}_{10}
\end{eqnarray}
One can now eliminate the diagonal terms from normalization to
transform the above relations to:
\begin{eqnarray}\label{RS-1'-expanded}
R^{00}_{10} + R^{01}_{10} + R^{10}_{00} + R^{11}_{00} &=&
R^{10}_{01} + R^{11}_{01} + R^{00}_{11} + R^{01}_{11}\\
R^{00}_{01} + R^{10}_{01} + R^{01}_{00} + R^{11}_{00} &=&
R^{00}_{11} + R^{10}_{11} + R^{01}_{10} + R^{11}_{10}
\end{eqnarray}
which are the relations given in \cite{Schutzstat}. In the next
section we will consider a two species model.
\section{A two species separable model}
As mentioned above the rates of the solvable two species models of
the type considered in this paper live on a $56$ dimensional
hyperplane. By exploring various regions of this plane, one can
find interesting two species models suitable for various
applications. In this section we explore a small region of this
plane by
imposing extra symmetry requirements on the model.\\
 We consider systems in which
 there is no driving force, i.e. systems which have the symmetry $ R_{\alpha,
 \beta}^{\mu, \nu} =  R_{\beta, \alpha
}^{\nu, \mu}$. As noted above (see equations (\ref{Ri},\ref{Si}),
for these models the ${\bf R}$ matrices and the
 ${\bf S}$ matrices lead to identical constraints.
Furthermore we restrict ourselves to those models in which an
individual species
 is neither created nor annihilated but only changes its label.
 Such models may be appropriate for description of voting processes
 or the spreading of epidemics. This means that we are setting
 $ R_{0,0}^{i,0} = R_{i,0}^{0,0} = R_{i,j}^{k,0} = R_{i,0}^{j,k} = 0 $
 In other words the number of Latin indices should be equal
 as subscripts and superscripts of $R$.
A rate such as $ R_{1,0}^{2,0}$ means that a voter with vote $1$
spontaneously (or due to the effect of environment) changes his or
her vote to vote $2$. (or a healthy individual $1$ gets infected
due to the interaction with the environment.) In a time interval
$dt$ two voters with different votes $1$ and $2$, pass each
other, without changing their votes, with probability $
R_{12}^{21}dt $. It may also happen that on this close contact
the voter $1$ changes his or her idea and switch to vote $2$.
This will happen with probability $R_{12}^{22}dt$. Note that the
voting processes that have been studied in the literature contain
only two species $+$ and $-$ with no vacant site. Here we have
also vacant sites, and the voters can move in free space and
interact with each other.\\
 Equations
(\ref{Ri}) now yields:
\begin{eqnarray}\label{RS-2}
  R_{10}^{10}+ R_{01}^{10} &=& R_{11}^{11} + R_{11}^{12}\hskip 2cm
  R_{10}^{20}+ R_{01}^{20}  =  R_{11}^{21} + R_{11}^{22}\\
  R_{10}^{10}+ R_{02}^{10} &=& R_{12}^{11} + R_{12}^{12}\hskip 2cm
  R_{10}^{20}+ R_{02}^{20}  =  R_{12}^{21} + R_{12}^{22}\\
  R_{20}^{10}+ R_{01}^{10} &=& R_{21}^{11} + R_{21}^{12}\hskip 2cm
  R_{20}^{20}+ R_{01}^{20}  =  R_{21}^{21} + R_{21}^{22}\\
  R_{20}^{10}+ R_{02}^{10} &=& R_{22}^{11} + R_{22}^{12}\hskip 2cm
  R_{20}^{20}+ R_{02}^{20}  =  R_{22}^{21} + R_{22}^{22}
  \end{eqnarray}
The diagonal terms like $ R_{10}^{10}, R_{20}^{20} \cdots $ are a
source of trouble, since they are minus the sum of a number of
rates and we would better get rid of them. To do so we proceed as
follows: We note that in each pair of equations above, the sum of
the right hand side terms adds up to zero, due to normalization.
Thus the sum of the left hand sides must also add up to zero. If
we do so and substitute for the diagonal terms their value from
the normalization (i.e: substitute $ R_{10}^{10} $ with
$-(R_{10}^{01}+ R_{10}^{02}+ R_{10}^{20})$, we find that the
first and the last pair of equations of the above set lead to
trivial identities.  We also find that the second and the third
pairs lead to one single identity, meaning that in each pair one
is redundant. Thus we keep this last identity and safely ignore
all the equations which contain diagonal terms. Therefore we are
actually dealing with $5$
independent equations relating positive rates, the final forms of which are:\\
\begin{eqnarray}\label{RS-2'}
  R_{10}^{20}+ R_{10}^{02} = R_{11}^{21} + R_{11}^{22}\ &&
  \
  R_{20}^{10}+ R_{20}^{01} = R_{22}^{12} + R_{22}^{11}\\
  R_{10}^{20}+ R_{02}^{20} = R_{12}^{21} + R_{12}^{22}\ &&
  \
  R_{20}^{10}+ R_{01}^{10} = R_{21}^{12} + R_{21}^{11}\\
  R_{01}^{10}+ R_{01}^{20}&=& R_{02}^{20} + R_{02}^{10}
\end{eqnarray}
These are the final conditions on the rates for this kind of two
species model.
\subsection{Equations of motion}
To obtain the equations of motion for the above two species
model, we use (\ref{linear1}) and obtain for a $d$ dimensional
rectangular lattice with unit vectors $ {\bf e}_r ; r = 1, \cdots
d $ and with the abbreviation $ \langle n_i({\bf x}) \rangle
\rightarrow n_i({\bf x}) $:
\begin{eqnarray}\label{linear1-example-1}
\frac{d}{dt} n_1({\bf x})&=&R_{01}^{10} \nabla^2 n_1({\bf x})-2 d
(R_{01}^{02}+R_{01}^{20}) n_1({\bf x})\cr &+& R_{02}^{10}\nabla^2
n_2({\bf x})+2d(R_{02}^{01}+ R_{02}^{10})n_2({\bf x}),
\end{eqnarray}
and
\begin{eqnarray}\label{linear1-example-2}
\frac{d}{dt} n_2({\bf x}) &=& R_{02}^{20}\nabla^2 n_2({\bf x}) -
2 d (R_{02}^{01} + R_{02}^{10}) n_2({\bf x})\cr &+&
R_{01}^{20}\nabla^2 n_{1}({\bf x}) + 2 d
(R_{01}^{02}+R_{01}^{20}) n_1({\bf x}).
\end{eqnarray}
where  $ \nabla^2 n({\bf x}) := \Big(\sum_{r=1}^{r=d} n({\bf
x}+{\bf e}_r )+n({\bf x}-{\bf e}_r ) - 2d n({\bf x})\Big)$ stands
for the discrete $d$ dimensional Laplacian.\\
All the terms in these equations can be understood intuitively.
For example a term like $ R_{02}^{10}\nabla^2 n_2({\bf x})$
measures the diffusion of particles of type 2 which change their
type or color to $1$ as they hop.\\ If one begins to write the
equations intuitively taking all the complex interactions into
account, one finds many other terms. But at the end they all
cancel out. For the description of the solution, it is convenient
to define new parameters:
\begin{eqnarray}\label{newparameters}
\gamma_1&:=& 2 d (R_{01}^{20}+R_{01}^{02}) \hskip 1cm  \gamma_{2}
=2 d (R_{02}^{10}+R_{02}^{01})\\
 \gamma&:=& \gamma_1 + \gamma_2\\
  D&:=& R_{01}^{10} + R_{01}^{20} = R_{02}^{20} + R_{02}^{10}\equiv 1\\
  D'&:=& R_{01}^{10} - R_{02}^{10} = R_{02}^{20} - R_{01}^{20},
\end{eqnarray}
where we have used the last relation of (\ref{RS-2'}) in the last
two relations and have rescaled time to set $ D = 1 $. These
parameters have obvious physical interpretations, $ \gamma_1 $
and $ \gamma_2$ respectively determine the overall tendency of
particles of type $1$ and $2$ to switch their types.  $ D $ is
the diffusion constant of the particles when we ignore their
types and finally  $ D'$ a kind of relative diffusion constant.
It measures the difference of diffusion constant for the particles
that do not change their type in hopping,
compared with those which do so.\\
Solving these coupled system of differential-difference equations
will give the distribution of both types of particles in space
and time. In principle it is possible to go to fourier space and
diagonalize the resulting matrix equation. However it is better
to proceed in a more physically transparent way by defining new
densities:
\begin{eqnarray}\label{newdensities}
  n({\bf x}) &=& n_1({\bf x}) + n_2 ({\bf x})\\
  \phi({\bf x}) &=& \gamma_1  n_1({\bf x}) - \gamma_2 n_2 ({\bf x})
\end{eqnarray}
Here $ n({\bf x})$ is the total density of particles (when we
ignore their types or colors) and $ \phi({\bf x}) $ is a weighted
difference of densities.\\
It is now a matter of simple algebra to use equations
(\ref{newparameters}) to arrive at the following equations for
these new densities:\\
\begin{eqnarray}\label{total}
\frac{\partial}{\partial t} n({\bf x}) &=& \nabla^2 n ({\bf x})
\end{eqnarray}
which means that if we ignore the color of particles, they perform
simple diffusion with diffusion constant $D$. We also obtain\\
\begin{eqnarray}\label{difference}
\frac{\partial}{\partial t} \phi({\bf x}) &=& D' \nabla^2
\phi({\bf x})+ D''\nabla^2 n({\bf x}) - \gamma \phi({\bf x})
\end{eqnarray}
where
\begin{equation}\label{C}
  D'':= R_{02}^{10}\gamma_1-R_{01}^{20}\gamma_2\equiv
      R_{02}^{10}R_{01}^{02}-R_{01}^{20}R_{02}^{01}
\end{equation}

The total number of particles of each species obey very simple
equations which are obtained by summing the above equations over
${\bf x}$. Denoting the number of particles of species $1$ and $2$
by $N_1$ and $N_2$ respectively, and noting that $ N_2 = N - N_1 $
where $ N$ is the total number of particles, we find from
(\ref{difference}) by summing $ \phi $ over all the sites:
\begin{equation}\label{N1equation}
\frac{d}{dt}N_1 = -\gamma_1 N_1 + \gamma_2 N_2 = -\gamma_1 N_1 +
\gamma_2 (N-N_1)
\end{equation}
the solution of which is:
\begin{eqnarray}\label{N1equation''}
N_1(t) &=& \frac{\gamma_2}{\gamma_1+\gamma_2}N +
e^{-(\gamma_1+\gamma_2)t}\big(N_1(0)-\frac{\gamma_2}{\gamma_1+\gamma_2}N\big)\\
N_2(t) &=& \frac{\gamma_1}{\gamma_1+\gamma_2}N +
e^{-(\gamma_1+\gamma_2)t}\big(N_2(0)-\frac{\gamma_1}{\gamma_1+\gamma_2}N\big)
\end{eqnarray}
We now discuss the solution of equations
(\ref{total}-\ref{difference}) which
determine the spatial distribution of particles in time.\\
Let us define the fourier transforms
\begin{equation}\label{generators}
  \bar{n}({\bf q},t):= \sum_{{\bf x}}e^{i{\bf q.x}} n({\bf x}) \hskip 2cm
  \bar{\phi}({\bf q},t):= \sum_{{\bf x}}e^{i{\bf q.x}} \phi({\bf
  x})
\end{equation}
where ${\bf q} = (q_1, q_2, \cdots q_d) $ and  for all $i$, $ q_i
\in [0,2\pi)$, from which we find:
\begin{equation}\label{generators'}
  n({\bf x}) = \int_{0}^{2\pi}e^{-i{\bf q.x}} \bar{n}({\bf q},t)\frac{d {\bf q}}{(2\pi)^d}\hskip 1cm
  \phi({\bf x}) = \int_{0}^{2\pi}e^{-i{\bf q.x}} \bar{\phi}({\bf
 q},t)\frac{d {\bf q}}{(2\pi)^d}
\end{equation}
With the definition:
\begin{equation}\label{Q}
  Q = 2\sum_{r=1}^{r=d} (\cos q_r -1)
\end{equation}
we find the equations of motion for these generating functions as:
\begin{eqnarray}\label{generatorsequation}
\dot{\bar{n}} &=& Q \bar{n}\\
\dot{\bar{\phi}} &=& Q (D' \bar{\phi} + D'' \bar{n}) - \gamma
\bar{\phi}
\end{eqnarray}

with the general solution:
\begin{eqnarray}\label{generatorsolution}
  \bar{n}(q,t) &=& \bar{n}(q,0) e^{Qt}\\
  \bar{\phi}(q,t) &=& e^{(QD'-\gamma)t}\big[\bar{\phi}(q,0)-
  \frac{QD''}{Q(1-D')+\gamma}\bar{n}(q,0)\big]\cr &+& e^{Qt}
  \big[ \frac{QD''}{Q(1-D')+\gamma}\bar{n}(q,0)\big]
\end{eqnarray}
Once the initial distributions of particles of each type is
known, these equations allow us to determine the distributions of
both types of particles in later times. If the particles are in a
finite volume, then the above solutions are still valid, except
that the momenta ${\bf q}$ will take discrete values.\\
The large scale behaviour of these densities is determined by
going to the limit of $ q\longrightarrow 0 $ where $ Q
\longrightarrow -|{\bf q}|^2  $. For illustration we consider two
simple examples.\\ \\
\textbf{Example 1:}  Let us assume that the particles (voters)
change their type (votes) only on encounter with other particles
(voters). In this case we have: $R_{10}^{20} = R_{10}^{02}=
R_{20}^{10} = R_{20}^{01} = 0 $. From (\ref{RS-2'}) we find that
the only non-zero parameters which remain are $ R_{12}^{21}=:P ,
R_{12}^{22}=R_{12}^{11}=: A $ and $ R_{10}^{01} =
R_{20}^{02}=:D$, subject to a relation: $ D = P + A $, where we
have introduced new simple labels for the rates. The labels $ D,
P $ and $ A$ stand respectively for " Diffusion ", " Pass" and "
Agreement":
\begin{eqnarray}\label{relations1}
&1\ \ 0& \leftrightarrow \ 0\ \ 1 \hskip 1cm  {\rm { with\ \ rate }}\ \ D\\
&2\ \ 0& \leftrightarrow \ 0\ \ 2 \hskip 1cm  {\rm { with\ \ rate }}\ \ D\\
&1\ \ 2& \leftrightarrow \ 2\ \ 1 \hskip 1cm  {\rm { with\ \ rate }}\ \ P\\
&1\ \ 2& \rightarrow \ 2\ \ 2 \hskip 1cm  {\rm { with\ \ rate }} \ \ A\\
&1\ \ 2& \rightarrow \ 1\ \ 1 \hskip 1cm  {\rm { with\ \ rate }}
\ \ A
\end{eqnarray}
In this case we find from (\ref{linear1-example-1} and
\ref{linear1-example-2}) that both types of particles diffuse
through each other without any interaction:
\begin{equation}\label{diffuse}
  \frac{\partial}{\partial t}n_1({\bf x}) = D \nabla^2 n_1(x),
  \hskip 2cm \frac{\partial}{\partial t}n_2({\bf x}) = D \nabla^2 n_2(x)
\end{equation}
However this is peculiar to one point functions and the equations
of two point functions will indeed be coupled to each other by
interaction parameters. This is an example of an observation
first made in \cite{afks} according to which some hamiltonians
may lead to the same set of equations for one point functions.
Here our hamiltonian is equivalent to a free hamiltonian as far
as the one point functions are concerned.\\ \\
\textbf{Example 2:} Let us now assume that one of the particles
say type $1$ does not change its type, or one kind of voters is
persistent in his vote, that is, $ R_{10}^{20}= R_{10}^{02} = 0
$. The first equation of (\ref{RS-2'}) then leads to $
R_{11}^{21} = R_{11}^{22} = 0$. In this case we find from
(\ref{newparameters}) that $D''=0 $ and $ D'= 1$. The relations
(\ref{RS-2'}) reduce to the following relations between the
remaining rates of reactions:
\begin{eqnarray}\label{RS-3}
  R_{02}^{20} &=& R_{12}^{21} + R_{12}^{22}\\
  R_{01}^{10} &=& R_{02}^{20} + R_{02}^{10}\\
  R_{20}^{10}+ R_{01}^{10} &=& R_{21}^{12} + R_{21}^{11}\\
  R_{20}^{10}+ R_{20}^{01} &=& R_{22}^{12} + R_{22}^{11}
\end{eqnarray}
For simplicity we first consider the large scale form of the
distribution functions. From (\ref{generatorsolution}) we obtain
\begin{eqnarray}\label{large-scale-2}
\bar{n}({\bf q},t) &=& \bar{n}({\bf q},0) e^{-|{\bf q}|^2 t}\\
\bar{\phi}({\bf q},t) &=& \bar{\phi}({\bf q},0) e^{(-|{\bf q}|^2
-\gamma)t}
\end{eqnarray}
Let us consider a situation at time $t=0$ where there are $N_2$
particles of type $2$ at the origin, i.e. $ n_2({\bf x}, 0) =
N_2\delta ( {\bf x})$ in a uniform see of particles of type $1$
of density $ \rho $, i.e. $ n_1({\bf x}, 0) = \rho$. Thus the
initial values of fourier transforms are:
\begin{equation}\label{initial}
  \bar{n}_1({\bf q},0) = \rho (2\pi)^d
  \delta({\bf q})\hskip 1cm \bar{n}_2({\bf q},0) = N_2
\end{equation} or
\begin{equation}\label{initial2}
  \bar{n}({\bf q},0) =  N_2 + \rho (2\pi)^d
  \delta({\bf q})\hskip 1cm \bar{\phi}({\bf q},0) = -\gamma_2 N_2.
\end{equation}

Inserting these into (\ref{large-scale-2}), we find:
\begin{eqnarray}\label{z1z2"}
\bar{n}(q,t) &=& (N_2 + \rho (2\pi)^d
  \delta({\bf q})) e^{-|{\bf q}|^2t} \equiv N_2 e^{-|{\bf q}|^2t} +  \rho (2\pi)^d
  \delta({\bf q}) \\
  \bar{\phi}(q,t) &=& -\gamma_2 N_2 e^{-(\gamma_2 + |{\bf q}|^2)t}.
\end{eqnarray}
Taking the inverse fourier transform and using the definition
(\ref{newdensities}) we find:
\begin{eqnarray}\label{inverse}
  n_1({\bf x},t) &=& \rho + \frac{N_2}{(4\pi t)^{\frac{d}{2}} } e^{\frac{-{|\bf x}|^2}{4t}}(1-e^{-\gamma_2 t})\\
  n_2({\bf x},t) &=& \frac{N_2}{(4\pi t)^{\frac{d}{2}}} e^{\frac{-{|\bf x}|^2}{4t}}e^{-\gamma_2 t}\\
 \end{eqnarray}
The initial number of particles of type $2$ diffuse and gradually
the particles of type $2$ turn into type $1$. At the end, no
particle of type $2$ remains.\\
We can also consider the small scale behavior of these
distributions. In this case the solution of the equations
(\ref{generatorsequation}) is given by modified Bessel functions.
We give below the solution for the initial distribution of one
particle of type 2 at the origin and a uniform distribution of
particles of type 1 at all other sites except the origin. That is
$( \ n_2({\bf x}, 0) = \delta_{{\bf x},0}\ \ {\rm or }\ \
\bar{n_2}({\bf q},0) = 1) \ $  and  ($ \ \ n_1({\bf x}) = \rho
(1-\delta_{{\bf x},0})\ \ {\rm or \ \ } , (\bar{n_1}({\bf q},0)=
\rho((2\pi)^d \delta(\bf{q})-1)$. For these initial conditions we
can find $ \bar{n}(\bar{q},t)$ and $ \bar{\phi}({\bf q},t)$ from
(\ref{generatorsolution}). The result is:
\begin{eqnarray}\label{bessel1}
\bar{n}({\bf q},t) &=& (1-\rho)e^{Qt} + (2\pi)^d \rho \delta({\bf
q})\\
\bar{\phi}({\bf q},t) &=& -\gamma_2 e^{-(Q-\gamma_2)t}
\end{eqnarray}
On the lattice we will find:
\begin{eqnarray}\label{final}
  n({\bf x},t) &=& \rho +
  (1-\rho)\frac{1}{(2\pi)^d}\int_{0}^{2\pi}e^{Qt-i{\bf q.x}}d{\bf
  q}\\
  \phi({\bf x},t) &=& -\gamma_2 e^{-\gamma_2 t} \frac{1}{(2\pi)^d}\int_{0}^{2\pi}e^{Qt-i{\bf q.x}}d{\bf
  q}.
\end{eqnarray}
These integrals can be written in terms of Modified Bessel
functions\\ $I_{x}(t) = \frac{1}{2\pi} \int_{0}^{2\pi} e^{-iqx +
\cos q t} dq $. Thus we find:
\begin{eqnarray}\label{final2}
  n_1({\bf x},t) &=& \rho + (
  1-\rho-e^{-\gamma_2t})e^{-2dt}\prod_{{\bf x}}I_{{\bf x}}(2t)\\
  n_2({\bf x},t) &=&  e^{-\gamma_2 t} e^{-2dt}\prod_{{\bf x}}I_{{\bf x}}(2t)
\end{eqnarray}

\section{Discussion}

We have obtained the general condition under which the hierarchy
of equations for $n-$ point functions of a multi-species reaction
diffusion system truncates at all orders, that is the equations of
motion of $n-$ point functions depend only on those of lower
correlation functions. Also under the above conditions the
equations of motion of two-point functions at different times
form a closed system. We have selected out a class of $2$ species
models which may be appropriate for the description of the spread
of an epidemic or as a voter model. This work can be extended in
several directions, even if one restricts oneself to the two
species models. First, one can study other two species models
which include coagulation, de-coagulation or birth and death
processes. Second, in the two species model one can change our
interpretation of $0$ particle as a hole and take it to be a real
particle. In this way one can adapt the rates for description of
other models. Third, one can study also the asymmetric models and
relax our simplifying assumption on the symmetry of rates, and
finally one can consider the similarity transformation
\cite{henkel, krebs, simon, amir} on the model to obtain other
exactly solvable models.

\section{Acknowledgement}
After this paper was submitted, I was informed by one of the
authors of ref. \cite{afks} that the question of decoupling of
equations of motion has already been addressed in their work and
the same equations (\ref{Ri} and \ref{Si}) have been obtained.
The considerations of \cite{afks} are however restricted to one
dimensional lattices. The result that the equations of two-point
functions at different times are closed is also new. On the other
hand the reader can find other interesting issues particularly
the so called gauge transformations and equivalent hamiltonians in
\cite{afks}. I would like to thank M. R. Rahimi-Tabar and H.
Arfaei for very useful discussions.

\end{document}